\documentclass[CEJP,DVI]{cej}

\usepackage{layout}
\usepackage{amsmath}
\usepackage{textcomp}

\usepackage{graphicx}
\usepackage{amssymb}

\newcommand{\nt}[1]{\lefteqn{#1}\slash}
\newcommand{\Tr}{\mathop{\mathrm{Tr}}}

\title{Factorization in the model of unstable particles with continuous masses}

\articletype{Research Article}

\author{Vladimir~I.~Kuksa\email{vkuksa47@mail.ru},
        Nikolay~I.~Volchanskiy\email{nikolay.volchanskiy@gmail.com}}

\institute{
     Research Institute of Physics,
     Southern Federal University,
     Prospekt Stachki, 194, 344090 Rostov-na-Donu, Russia
          }

\abstract{We study processes with unstable particles in intermediate time-like states. It is shown that the amplitudes squared of such processes factor exactly in the framework of the model of unstable particles with continuous masses. Decay widths and cross sections can then be represented in a universal factorized form for an arbitrary set of interacting particles. This exact factorization is caused by specific structure of propagators in the model. We formulate the factorization method and perform a phenomenological analysis of the factorization effects. The factorization method considerably simplifies calculations while leading to compact and reasonable results.}


\begin{document}

\maketitle


\section{\label{sec:1}Introduction}


Field-theory description of unstable particles (UP's) with a large width runs into some problems which have been under considerable discussion for many decades (see \cite{1958PhRv..112..283M, 1959PhRv..115.1079M, Levy:443619, 1960AnPhy...9..169S, 1961PhRv..121..350J, Veltman1963186, Burgers:189859, 1988ZPhyC..40..141B, 1991PhLB..267..240S, 1991PhLB..262..113S, 1991PhLB..259..373W, 1996PhRvD..53.2128P, 1998PhRvD..58k3010P, 2000NuPhB.581...91B, 2001PhRvL..86..389G, 2002PhRvD..65h5001G, 2003quant.ph.12178K} and references therein). These problems are both conceptual and computational in character and arise due to the fact that UP's lie somewhat outside the traditional formulation of quantum theory \cite{Veltman1963186, 2000NuPhB.581...91B, 2003quant.ph.12178K}. An unstable particle with a large width cannot be treated as an asymptotic initial or final state. So, a UP is usually described in an intermediate state with the help of the dressed propagator or the $S$-matrix with a complex pole. However, the application of the Dyson procedure leads to a deviation from the scheme of fixed-order calculations and to some problems with gauge cancellation \cite{Altarelli:1996ww, 1998NCimR..21i...1M, bardin1999standard}. In order to overcome these obstacles, some approximation schemes and methods have been worked out. For instance, double-pole approximation (DPA) \cite{2000hep.ph....5309G, 1999NuPhB.548....3B, 1995NuPhB.440...95D, 1998NuPhB.519...39D}, pinch-technique method \cite{1995PhRvL..75.3060P, PhysRevD.66.111901}, complex-mass scheme (CMS) \cite{2005NuPhB.724..247D, 2005PhLB..612..223D}, semi-analytical approximation \cite{bardin1999standard}, narrow-width approximation (NWA) \cite{bardin1999standard, 2007PhRvL..99k1601B}, convolution method (CM) \cite{Altarelli2001125, 2005PhRvD..72e5018B, 1992PhRvD..45..982K, 1992ZPhyC..56..273U, 1992PhRvD..46.2982U, 2000JPhG...26..387K}, etc. At the same time, alternative ways to deal with UPs were developed, such as effective theories of UP's \cite{Chapovsky2002257, 2004PhRvL..93a1602B}, modified perturbation theory \cite{2002PhLB..531..225N, 2001EPJC...19..441N} and models with continuous mass spectrum \cite{1961AnPhy..16..158G, 1965AnPhy..34..161L, lukierski1, lukierski2, lukierski3, lukierski4, 2009IJMPA..24.1185K}.

The properties of UP have also been a subject of discussion in the literature. In particular, the assumption that the decay of unstable particle $R$ proceeds independently of its production remains of interest \cite{jackson1964remarks, Pisut1968325, lichard1999photon}. Formally, this effect is expressed quantitatively as the factorization of a cross section or decay width \cite{lichard1999photon}. The processes of the type $ab\to Rx\to cdx$ were considered in \cite{lichard1999photon}. It was shown that in the framework of the traditional approach the factorization is always valid for a scalar $R$ and does not take place for a vector or spinor $R$. The effect of the factorization is usually considered as related to NWA \cite{bardin1999standard}, which is based on five assumptions \cite{2007PhRvL..99k1601B}.

In this work, we derive the continuous-mass propagators and analyse the factorization effects in the framework of a model that describes UP's as the generalized fields with continuous mass spectrum introduced in \cite{1961AnPhy..16..158G, 1965AnPhy..34..161L}. The formalism of the generalized fields was developed further in \cite{lukierski1, lukierski2, lukierski3}.

In our previous papers \cite{2006PhLB..633..545K, 2008IJMPA..23.4509K}, it was shown by straightforward calculations that expressions for the width of the decay $a\to bR\to bcd$ and for the cross section of the process $ab\to R\to cd$ factor exactly for scalar, vector and spinor $R$. The exact factorization is caused by the specific structure of the propagators of UP's with a continuous mass. The factorization method based on these results was developed in \cite{2009PAN....72.1063K, 2010IJMPA..25.2049K}. It has been applied to some processes that are studied experimentally \cite{2009IJMPA..24.1185K, 2008IJMPA..23.4125K, 2009IJMPA..24.5765K, 2010PAN....73.1622K} or could be studied in future experiments \cite{2011MPLA...26.1075P, 2012IJMPA..2750072K}. In all cases, it was found that the difference between continuous-mass--model predictions and standard calculations does not exceed 0.1~\%, much less than experimental uncertainties.

Note that the discussion in \cite{2006PhLB..633..545K, 2008IJMPA..23.4509K, 2009PAN....72.1063K, 2010IJMPA..25.2049K} was limited to some of the simplest tree diagrams. Here, we consider a more general case---vertices having an arbitrary (loop) structure and UP carrying an arbitrary spin $j_R$. It is shown that the amplitude squared can be exactly factorized upon integration over the phase space of the final states. On the basis of this analysis, the factorization method \cite{2010IJMPA..25.2049K} receives further development. 

The paper is organized as follows. In Sec.~\ref{sec:2}, we derive expressions for the model propagators and systematically study the factorization of the amplitudes squared for the processes with UP in an intermediate time-like state. Universal factorized formulae for decay widths and cross sections are derived in Sec.~\ref{sec:3}. The factorization approach is applied to the processes with several consequent decays of UP's in Sec.~\ref{sec:4}, where the factorization method is formulated in its most general form. In Sec.~\ref{sec:4}, we also discuss some methodological and phenomenological aspects of the factorization. Appendices give details of the mathematical proofs.


\section{\label{sec:2}Factorization effects in the model of unstable particles with continuous masses}


In this section, we consider the structure of the amplitudes for the processes $X_I \to R \to X_F$ ($X_I$ and $X_F$ are arbitrary sets of initial and final states; $R$ stands for an unstable particle in an intermediate state transferring a time-like momentum). We show that a special form of the propagators of unstable fields leads to the factorization of transition probabilities. In contrast to the traditional treatment (the narrow-width approximation), the approach under consideration provides the exact factorization for any type of UP with arbitrary width. This effect makes it possible to represent decay widths and cross sections in a universal factorized form.

The model under consideration is based upon uncertainty relation $\delta m \cdot \delta \tau \sim 1 $ \cite{1999JMP....40.1237K, 2002tqm..conf...69B, 1958PhRv..112..283M, 1961PhRv..121..350J}. In our model the notion of mass indefiniteness following from the mass-lifetime uncertainty relation is formalized by using generalized free fields \cite{1961AnPhy..16..158G, 1965AnPhy..34..161L}
\begin{equation}\label{eq:GFF}
	\phi(x) = \int w(\mu) \phi(x,\mu) d \mu,
\end{equation}
where $\phi(x,\mu)$ is a canonical free field with a continuous mass $\mu$, and $w(\mu)$ is a weight function. The generalized fields are treated as asymptotic states, which leads to convolution formulae for observables (see Sec. \ref{sec:4} and \ref{sec:5}). The mass indefiniteness is linked to the contributions of vacuum fluctuations to UP mass \cite{2009IJMPA..24.1185K, 2009PAN....72.1063K}.

The canonical commutation relation is generalized straightforwardly supposing that acts of creation/annihilation of UP with different values of mass parameter $\mu$ are independent:
\begin{equation}\label{eq:ccr}
	\left[ \bar\phi^+ (\mathbf{k}, \mu), \phi^- (\mathbf{p}, \mu')\right]
	= \delta (\mathbf{k}-\mathbf{p}) \delta (\mu-\mu').
\end{equation}

Using the commutation relation, we can calculate the propagator of a scalar generalized field. The propagator is given by a sum of the fixed-mass Green's functions similar to the K\"{a}ll\'{e}n--Lehmann spectral representation \cite{2009IJMPA..24.1185K}:
\begin{equation}\label{eq:D0}
	D(q^2)
	= i \int d x e^{-iqx} \langle 0 | \hat{T} \phi(x) \phi(0) |0 \rangle
	= i \int_{\mu_0}^{+\infty} \frac{\rho(\mu^2) d\mu^2}{q^2 -\mu^2 +i\epsilon},
\end{equation}
where $\mu_0$ is a decay threshold and $\rho(\mu^2) = \lvert w(\mu) \rvert^2$ is a probability density of the mass squared.

Using the Sokhotski--Plemelj formula
\begin{equation}\label{}
    \frac{1}{x \pm i\epsilon} = \mp i\pi\delta(x) + \mathcal{P} \frac1{x},
\end{equation}
we can obtain the real part of the propagator \eqref{eq:D0}:
\begin{equation}\label{eq:opt0}
    \Re D(q^2) = \frac{1}{2} \left[ D(q^2) + D^*(q^2)\right] = \pi \rho(q^2).
\end{equation}
In a particular case of the propagator defined in the conventional Dyson way,
\begin{equation}\label{}
    D(q^2) = \frac{i}{q^2 -M^2(q) +i\Im\Pi(q)},
\end{equation}
it follows from Eq.~\eqref{eq:opt0} that the spectral density takes the form
\begin{equation}\label{}
    \rho(q^2)
    =
    \frac{1}{\pi} \frac{\Im \Pi(q)}{ \left[q^2-M^2(q)\right]^2+\left[\Im \Pi(q)\right]^2}.
\end{equation}

Now, let us consider the spectral-like representation for the propagator of a vector field
\begin{equation}\label{eq:D1}
    D_{\mu\nu}(q^2)
    =
     i \int_{\mu_0}^{+\infty}
     \left( g_{\mu\nu}- \frac{q_\mu q_\nu}{\mu^2} \right) \frac{\rho(\mu^2) d\mu^2}{q^2 -\mu^2 +i\epsilon}.
\end{equation}
The real part of the above propagator is as follows:
\begin{equation}\label{eq:opt2}
    \Re D_{\mu\nu}(q^2)
    =
     \pi \left( g_{\mu\nu}- \frac{q_\mu q_\nu}{q^2} \right) \rho(q^2).
\end{equation}
Comparing Eqs.~\eqref{eq:opt0} and \eqref{eq:opt2}, we find that the propagators of scalar and vector fields in the continuous-mass model can be written as
\begin{equation}\label{2.1}
D(q) = \frac{i}{P_0(q)},
\qquad
D_{\mu\nu}(q) = -i\frac{g_{\mu\nu}-q_{\mu}q_{\nu}/q^2}{P_1(q)}.
\end{equation}

In the general case, the propagators of the fields carrying integer spins $j = \ell$ and half-integer ones $j = \ell + 1/2$ ($\ell = 1,2,\dots$) can be written, respectively, as
\begin{equation}\label{2.1a}
D_{\bar\mu\bar\nu}(q) = -\frac{i}{P_j(q)} P^{(\ell)}_{\bar\mu\bar\nu}(q),
\qquad
\hat D(q) = -i\frac{\nt{q}+q}{P_{1/2}(q)}
\qquad
\hat D_{\bar\mu\bar\nu}(q) = -i\frac{\nt{q}+q}{P_j(q)} P^{(\ell+1/2)}_{\bar\mu\bar\nu}(q).
\end{equation}
Here, $\bar\mu = \mu_1\mu_2\dots\mu_\ell$, $\bar\nu = \nu_1\nu_2\dots\nu_\ell$ are multi-indices; $P^{(j)}_{\bar\mu\bar\nu}$ are the spin-$j$ projection operators defined in Appendix \ref{app:a3}.

In Eqs.~\eqref{2.1} and \eqref{2.1a} the denominators $P_j(q)$ can be taken in various forms (pole, Breit--Wigner and other definitions). It is essential that the effect of factorization does not depend on the form of the denominators $P(q)$, but depends crucially on the tensor-spinor structure of the numerators in \eqref{2.1} and \eqref{2.1a}. The propagators \eqref{2.1} and \eqref{2.1a} with a continuous mass parameter $q$ result in the exact factorization, while conventional field-theory expressions with a constant mass $M$ lead to an approximate factorization in the narrow-width limit. We note that the factorization occurs in any model leading to the spectral representation \eqref{eq:D0} and \eqref{eq:D1} of the propagators. Also, it should be noted that there is no unique definition of the UP propagators in the literature. For example, the following expressions are used instead of $P_{\mu\nu}^{(1)}(q)$: $g_{\mu\nu} - q_\mu q_\nu / (m-i \Gamma/2)^2$ \cite{Altarelli2001125}, $g_{\mu\nu} - q_\mu q_\nu / (m^2-i m \Gamma)$ \cite{2005PhRvD..72e5018B}, and $P_{\mu\nu}^{(1)}(q) a(q^2) + q_\mu q_\nu / q^2 b(q^2)$ \cite{1993PhRvL..70.3193S}.

It should be emphasized that the structures \eqref{2.1} and \eqref{2.1a} are not constrained by the choice of the gauge. The model under consideration is not a gauge one, because it describes effective unstable fields (see \cite{2009IJMPA..24.1185K} for details). We also note that the difference between the model and field-theory functions is given by the value $(q^2-M^2)/M^2$, which is small in the vicinity of a resonance. Moreover, it is strongly suppressed by additional small factors \cite{2010IJMPA..25.2049K}. So, the approach discussed can be treated as some close approximation to the standard one, that is, it gives an analytical alternative to NWA (see Sec.~\ref{sec:4} and \cite{2010IJMPA..25.2049K}).

The factorization can be illustrated diagrammatically by cutting an internal line that stands for UP with a time-like momentum, if such an operation disconnects the diagram into two independent fragments. So, in the framework of the phenomenological model under consideration we need to include UP in initial and final states. The polarization density matrices of vector, spinor, and vector-spinor UPs are defined as the following spin sums
\cite{2009PAN....72.1063K, 2010IJMPA..25.2049K}:
\begin{equation}\label{2.2}
\begin{aligned}
\Pi_{\mu\nu}(q) &{}= \sum_{a=1}^{3} e^a_\mu(\mathbf{q})\dot{e}^a_\nu(\mathbf{q})
 = -g_{\mu\nu}+\frac{q_\mu q_\nu}{q^2},
\\
\hat\Pi(q)
&{} = \sum_{a=1}^{2}u^{a,\mp}(\mathbf{q})\bar{u}^{a,\pm}(\mathbf{q})
= \frac{\nt{q} \pm q}{2q^0},
\\
\hat\Pi_{\mu\nu}(q) &{}=
-\frac{\nt{q} \pm q}{2q^0} \biggl[ g_{\mu\nu}-\frac13 \gamma_\mu \gamma_\nu
- \frac{\gamma_\mu q_\nu -\gamma_\nu q_\mu}{3q} - \frac23 \frac{q_\mu q_\nu}{q^2} \biggr].
\end{aligned}
\end{equation}
For higher spins $j \geqslant 1$ the operators \eqref{2.2} become
\begin{equation}\label{2.2a}
\Pi_{\bar\mu\bar\nu}(q) = -P^{(\ell)}_{\bar\mu\bar\nu}(q),
\qquad
\hat \Pi_{\bar\mu\bar\nu}(q) = -\frac{\nt{q} \pm q}{2q^0} P^{(\ell+1/2)}_{\bar\mu\bar\nu}(q).
\end{equation}

Further, we show that the similarity of the propagators \eqref{2.1}, \eqref{2.1a} and the polarization matrices \eqref{2.2}, \eqref{2.2a} leads to the exact factorization of the cross sections and decay widths in a few of the  simplest special cases. In Appendix \ref{app:a}, we generalize the result to any interaction mediated by UP with arbitrarily high spin.

\begin{figure}
    \center\includegraphics{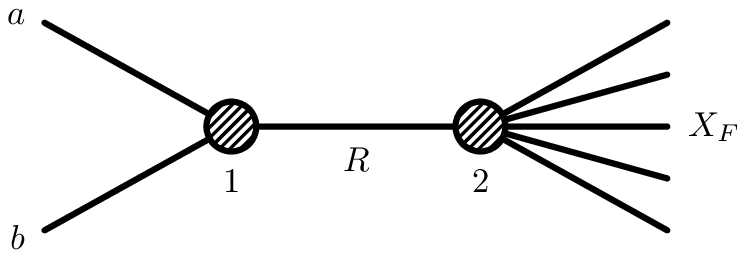}
    \caption{\label{fig:Born4}The scattering $ab\to R\to X_F$.}
\end{figure}

To give an example of the factorization, let us consider first the simplest two-particle scattering $ab \to R \to cd$, where $R$ is UP with a large width (see Fig.~\ref{fig:Born4}). In the case of a spinor UP we can prove by simple calculation that the amplitude squared of the process $\phi_1\psi_1\to \psi_R\to\phi_2\psi_2$ ($\phi_a$ and $\psi_a$ are scalar and fermion particles, respectively) can be written in a factorized form
\begin{equation}\label{2.3}
|\mathcal{M}|^2 =
\frac{4(q^0)^2}{|P_R(q)|^2} |\mathcal{M}_1|^2\cdot
|\mathcal{M}_2|^2.
\end{equation}
Here, $\mathcal{M}_1$ and $\mathcal{M}_2$ are the amplitudes of the processes $\psi_R\to \phi_1\psi_1$ and $\psi_R\to \phi_2\psi_2$, respectively. The amplitudes squared $|\mathcal{M}|^2$ and $|\mathcal{M}_i|^2$ contain traces of $4 \times 4$ spinor matrices and the factorization in Eq.~\eqref{2.3} is caused by the factorization of the trace
\begin{equation}
\Tr\Bigl[(\nt{k}_1+m_2)(\nt{q}+q)(\nt{p}_1+m_1)(\nt{q}+q)\Bigr]
= 2 \Tr\Bigl[(\nt{k}_1+m_2)(\nt{q}+q)\Bigr]\cdot
\Tr\Bigl[(\nt{p}_1+m_1)(\nt{q}+q)\Bigr],
\label{2.4}
\end{equation}
where $q=p_1+p_2$; $p_1$ and $k_1$ are the momenta of spinor particles in the initial and final states; $m_{1,2}$ are their masses.

By direct calculations it can be checked that the factorized relation \eqref{2.3} is not valid for the process $\phi_1 V_1 \to V_R \to \phi_2V_2$, where $\phi$ denotes scalar fields and $V$ are vector fields. However, the factorization is restored upon integrating over the phase space of the final states ($\phi_2, V_2$):
\begin{equation}\label{2.5}
J(|\mathcal{M}|^2)=
\frac{4(q^0)^2}{|P_V(q)|^2} |\mathcal{M}_1|^2 \cdot J(|\mathcal{M}_2|^2).
\end{equation}
Here, the operator $J(A)$ denotes the integration over the phase space of the final momenta $k_1$ and $k_2$ (see Eq.~\eqref{app:int1} in Appendix \ref{app:a}). The relation \eqref{2.5} can be verified by direct calculations, however, we prove it in a more general way. In the case under consideration, the amplitude squared $|\mathcal{M}(p,q,k)|^2$ is given by
\begin{equation}\label{2.6}
\begin{aligned}
|\mathcal{M}(p,q,k)|^2&=\frac{12(q^0)^2}{|P_R(q)|^2}M_{(1)}^{\mu\nu}(p,q)M^{(2)}_{\mu\nu}(k,q),
\\
|\mathcal{M}_1(p,q)|^2 &\equiv M^\mu_{(1)\mu}(p,q) =\frac{g^2_1}{24q^0p^0_a p^0_b}
\eta_{\mu\nu}(p_a)\eta^{\mu\nu}(q),
\\
|\mathcal{M}_2(k,q)|^2 &\equiv M^\mu_{(2)\mu}(k,q)
= \frac{g^2_2}{24q^0k^0_1k^0_2} \eta_{\mu\nu}(k_1)\eta^{\mu\nu}(q).
\end{aligned}
\end{equation}
Here, $g_k$ ($k=1,\,2$) are coupling constants; $\eta_{\mu\nu}(q) = -g_{\mu\nu} + q_{\mu} q_{\nu} / q^2$. Integrating $M_{\mu\nu}^{(2)}(k,q)$ over the final momenta $k_1$ and $k_2$ gives
\begin{equation}\label{2.7}
J\bigl[M_{(2)}^{\mu\nu}(k,q)\bigr] \equiv
\int d\mathbf{k}_1 d\mathbf{k}_2 \delta(q-k_1-k_2) M_{(2)}^{\mu\nu}(k,q)
= c_1(q) g^{\mu\nu} + c_2(q) q^{\mu} q^{\nu}.
\end{equation}
Multiplying Eq.~\eqref{2.7} by $\eta^{\mu\nu}(q)$ we get
\begin{equation}\label{2.8}
c_1(q) = \frac13 J\bigl[M^{\mu}_{(2)\mu}(k,q)\bigr]
=\frac13 J\bigl[|\mathcal{M}_2(k,q)|^2\bigr].
\end{equation}
It is immediately seen that Eq.~\eqref{2.5} follows from Eqs.~\eqref{2.6}, \eqref{2.7}, and \eqref{2.8}. Generalizing this result to the case of an arbitrary multi-particle final state and arbitrary structure of the vertices (which might include loop contributions) is considered in Appendix \ref{app:a}. In this general case (see Appendix~\ref{app:a}), Eq.~\eqref{2.5} becomes
\begin{equation}\label{2.9}
J\bigl[|\mathcal{M}(p,k,q)|^2\bigr] =
\frac{(2j_R+1)(2q^0)^2}{(2j_a+1)(2j_b+1)|P_R(q)|^2}
|\mathcal{M}_1(p,q)|^2 J\bigl[|\mathcal{M}_2(k,q)|^2\bigr].
\end{equation}
Here, $j_{a,b}$ are the spins of the initial particles, $j_R$ is an integer spin of the resonance, and $J$ denotes the integration over the multi-particle phase space. It should be noted that a stronger factorized relation is valid for a broad class of interactions of UPs carrying half-integer spins (including gauge and Fermi's interactions of fermions). In this case, the same factorization \eqref{2.9} of the amplitude squared exists even without the integration $J$ over the phase space of the final particles.


\section{\label{sec:3}Factorized formulae for the observables}


For any kinds of particles in the initial $X_I= (a, b)$, final $X_F$, and intermediate $R$ states the cross section of the scattering $ab \to R \to X_F$ (Fig.~\ref{fig:Born4}) can be written in a factorized form (see Appendix \ref{app:b}):
\begin{equation}\label{3.2}
\sigma (ab \to R \to X_F)
=
\frac{16\pi L_R}{L_a L_b \bar{\lambda}^2(m^2_a,m^2_b;s)} \frac1{|P_R(s)|^2}
\Gamma(R(s)\to ab) \Gamma(R(s) \to X_F).
\end{equation}
Here, $L_i=2j_i+1$; $s = q^2$; $R(s)$ is UP with a continuous mass $\sqrt{s}$; $\bar\lambda(m^2_a,m^2_b;s)$ is an analogue of the K\"{a}ll\'{e}n function
\begin{equation}\label{Kallen}
    \bar\lambda(p^2_a,p^2_b;s)
     = \left[ 1 -2 \frac{m^2_a + m^2_b}{s} + \frac{(m^2_a - m^2_b)^2}{s^2} \right]^{1/2}.
\end{equation}

The cross section \eqref{3.2} does not depend explicitly on the spins of the final states. Therefore, it can be summed over the final channels $R(s)\to X_F$
\begin{equation}\label{3.3}
 \sigma_\mathrm{inclusive}(ab\to R)=
\frac{16\pi^2 L_R \Gamma(R(s)\to
  ab)}{L_a L_b \sqrt{s}\bar{\lambda}^2(m^2_a,m^2_b;s)}\rho_R(s),
\end{equation}
where the function
\begin{equation}\label{distr}
    \rho_R(s)= \frac{\sqrt{s}\Gamma^\mathrm{tot}_R(s)}{\pi|P_R(s)|^2}
\end{equation}
is interpreted in the model of UP with continuous mass as probability density of UP mass \cite{2009IJMPA..24.1185K}.

\begin{figure}
    \center\includegraphics{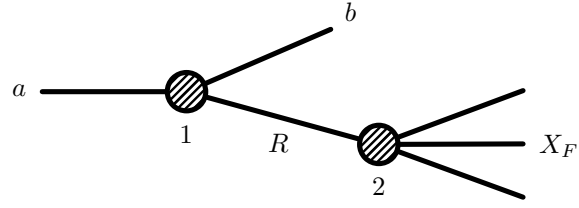}
    \caption{\label{fig:Born4a}Factorization in $a\to bX_F$ decay diagram.}
\end{figure}

Now, we consider the decay process of type $a\to bR\to bX_F$, where $R$ is UP with an arbitrary spin, $a$ and $b$ are quasi-stable particles with negligible width, and $X_F$ is the set of final states (see Fig.~\ref{fig:Born4a}).

For the decay process $a \to b R \to b X_F$ the relation similar to Eq.~\eqref{2.9} holds (see Appendix~\ref{app:a}):
\begin{equation}\label{3.4}
J\bigl[|\mathcal{M}(p,k,q)|^2\bigr]
=\frac{4(q^0)^2}{|P_R(q)|^2}
|\mathcal{M}_1(p,q)|^2 J\bigl[|\mathcal{M}_2(k,q)|^2\bigr],
\end{equation}
where $\mathcal{M}_1(p,q) = \mathcal{M}_1(a \to b R)$, $\mathcal{M}_2(k,q) = \mathcal{M}_2(R \to X_F)$. We can then calculate the width of the decay (see Appendix \ref{app:b}):
\begin{equation}\label{3.5}
\Gamma(a\to bX_F)=\int \Gamma(a\to bR(q))\frac{q\Gamma(R(q)\to
X_F)}{\pi|P_R(q)|^2}dq^2.
\end{equation}
In the above equation, the domain of integration is constrained by the kinematics of the reaction. Summing over all the decay channels of UP we get
\begin{equation}\label{3.6}
\Gamma(a\to bR)=\int \Gamma(a\to bR(q))\rho_R(q^2)dq^2,
\end{equation}
where the function $\rho_R(q^2)$ is defined above in Eq.~\eqref{distr}. Note that the factorization \eqref{3.2} and \eqref{3.5} is exact within the framework of the continuous-mass model and approximate in the standard treatment (the narrow-width approximation and convolution method).

The factorization of the cross section \eqref{3.2} and width \eqref{3.5} is a direct consequence of the factorization of the amplitude squared \eqref{2.9}. In \cite{2006PhLB..633..545K, 2008IJMPA..23.4509K}, the formulae \eqref{3.2} and \eqref{3.5} were obtained in some special cases (for specific tree processes in the case of the lowest-spin UP, $j_R=0,\,1/2,\,1$). In Appendices \ref{app:a} and \ref{app:b} these relations are derived in the most general case---for any spin of UP, any set of initial and final particles, and any structure of the vertices involved.

It should be noted that for scalar and fermion UP ($j_R=0,1/2,3/2,\dots $) the differential cross sections and decay widths also take a factorized form (see Appendix \ref{app:a}).


\section{\label{sec:4}Factorization method in the model of UP's with continuous mass}


The method is based on the exact factorization \eqref{3.2}, \eqref{3.5} of the simplest processes with UP in an intermediate state that were considered in Sec.~\ref{sec:3} (see also \cite{2009PAN....72.1063K, 2010IJMPA..25.2049K}). It can be applied to complicated decay chains and scattering processes that can be reduced to a tree constructed from the basic subdiagrams depicted in Figs.~\ref{fig:Born4} and \ref{fig:Born4a} (loop contributions are limited to the vertex blobs). Further, we consider some examples of such processes with UP's in time-like intermediate states.

\subsection{Process $ab \to R_1 \to cR_2\to cX_F$}

\begin{figure}
    \center\includegraphics{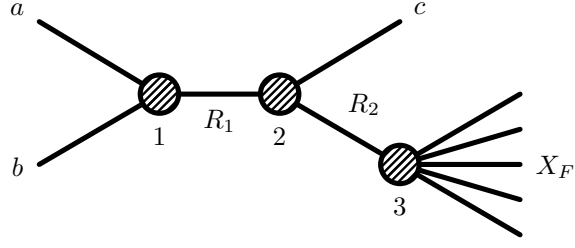}
    \caption{\label{fig:Born4b}Factorization in $ab\to cX_F$ scattering-decay diagram.}
\end{figure}

The cross section of this process (Fig.~\ref{fig:Born4b}) is a combination of the expressions \eqref{3.2} and \eqref{3.5}:
\begin{equation}\label{4.1}
\sigma(ab \to cX_F)
 =
 \frac{16L_{R_1}}{L_a L_b\bar{\lambda}^2(m_a,m_b;\sqrt{s})}
 \frac{\Gamma^{ab}_{R_1}(s)}{|P_{R_1}(s)|^2}
 \int_{q^2_1}^{q^2_2}\Gamma(R_1(s)\to c R_2(q))
 \frac{q \Gamma^{X_F}_{R_2}(q)}{|P_{R_2}(q)|^2} dq^2,
\end{equation}
where $L_i=2j_i+1$ and $\Gamma^B_A(q)=\Gamma(A(q)\to B)$. It should be noted that the factorization reduces the number of independent kinematic variables to be integrated over. For example, in the general case of the $2 \to 3$ process there are $N = 5$ variables that uniquely specify a point in the phase space, four of which have to be integrated over \cite{1969PhRv..185.1865K}. Some of these variables can be easily integrated out if the process possesses specific symmetry. In the framework of the approach suggested, the number of variables being integrated is always $N' = 1$ (the variable $q^2$ in Eq.~\eqref{4.1}).


\subsection{Decay chain}

\begin{figure}
    \center\includegraphics{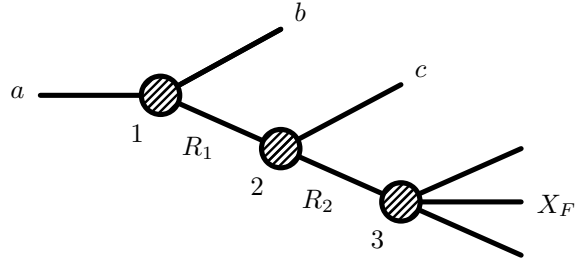}
    \caption{\label{fig:Born5}Factorization in $a\to bcX_F$ decay diagram.}
\end{figure}

The width of the decay chain $a\to bR_1\to bcR_2\to bcX_F$ depicted in Fig.~\ref{fig:Born5} is a direct consequence of Eq.~\eqref{3.5}:
\begin{equation}\label{4.2}
\Gamma(a\to bcX_F) = \int_ {q^2_1}^{q^2_2}
\frac{q \Gamma(a \rightarrow b R_1(q))}{\pi\vert P_{R_1}(q)\vert^2}
\int_{g^2_1}^{g^2_2}\Gamma(R_1(q)\to cR_2(g))
\frac{g \Gamma(R_2(g)\rightarrow X_F)}{\pi\vert P_{R_2}(g)\vert ^2} dg^2 dq^2.
\end{equation}
Note that in the general case the number of kinematic variables which uniquely specify a point in the four-particle phase space is $N=5$ \cite{1969PhRv..185.1865K}, while the model under consideration leaves $N'=2$ (invariants $q^2$ and $g^2$).


\subsection{Decay of the resonance produced in $t$-channel scattering}

\begin{figure}
    \center\includegraphics{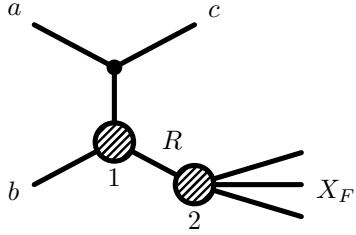}
    \caption{\label{fig:Born6}Factorization in $ab\to cR\to cX_F$ process.}
\end{figure}

The cross section of the process $ab\to cR\to cX_F$ (Fig.~\ref{fig:Born6}) is given by a convolution of the cross section $\sigma(ab \to cR)$ and the width $\Gamma(R\to X_F)$:
\begin{equation}\label{4.3}
\sigma(ab\to cX_F)=\int_{q_1^2}^{q_2^2}\sigma(ab\to
cR(q))\frac{q\Gamma(R(q)\to X_F)}{\pi\vert P_{R}(g)\vert ^2}\,dq^2.
\end{equation}
This formula can be applied to the description of the processes $e^+e^-\to \gamma Z\to \gamma f\bar{f}$ \cite{2010PAN....73.1622K} and $eN\to e\Delta \to e\pi N$ \cite{2010IJMPA..25.2049K}.


\subsection{Resonance-pair production $ab\to R_1 R_2\to X_1 X_2$}

\begin{figure}
    \center\includegraphics{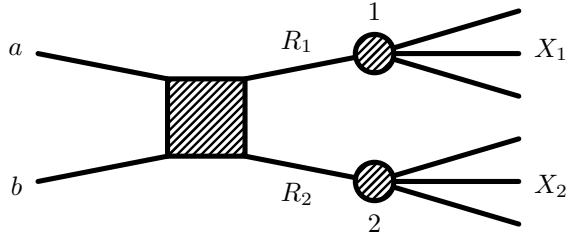}
    \caption{\label{fig:Born7}$R_1 R_2$-pair production process.}
\end{figure}

Now, we consider the process of boson-pair production depicted in Fig.~\ref{fig:Born7} (e.g. four-fermion production in the double-pole approximation). Applying the model to the process $e^+e^-\to R_1R_2$ directly or using the factorization method for the whole process $ab\to R_1 R_2\to X_1 X_2$ leads to the following expression for the inclusive cross section at the tree level \cite{2009IJMPA..24.1185K}:
\begin{equation}\label{4.4}
\sigma_\mathrm{tree}(e^+e^- \to R_1 R_2)
=
\int \rho_1(m_1) \rho_2(m_2)
\sigma_\mathrm{tree}[e^+e^- \to R_1(m_1)R_2(m_2)] dm^2_1 dm^2_2,
\end{equation}
where $\sigma^\mathrm{tr}[e^+e^-\to R_1(m_1)R_2(m_2)]$ is a cross section for the case of the fixed boson masses $m_1$ and $m_2$, while the probability density function of a mass $\rho(m)$ is defined by the expression
\begin{equation}\label{4.5}
\rho_R(m)=\frac{1}{\pi} \frac{m \Gamma^\mathrm{tot}_R(m)}{(m^2-M^2_R)^2+(m \Gamma^\mathrm{tot}_R(m))^2}.
\end{equation}
These expressions can be applied to the processes $e^+e^-\to ZZ, W^+W^-$ \cite{2008IJMPA..23.4125K, 2009IJMPA..24.5765K} and $e^+e^-\to ZH$ \cite{2010PAN....73.1622K}. To describe exclusive processes $e^+e^-\to R_1R_2\to f_i\bar{f}_{i'} f_k\bar{f}_{k'}$, we have to substitute the decay width $\Gamma_R^i=\Gamma(R\to f_i\bar{f}_{i'})$ for the total width $\Gamma^\mathrm{tot}_R(m)$ in the numerator of the righthand side of Eq.~\eqref{4.5} (double-pole approximation).

Using the factorization method one can describe complicated decay-chain and scattering processes in a simple way. The same results could be obtained as approximations within the framework of the standard treatment, such as the narrow-width approximation \cite{bardin1999standard, 2007PhRvL..99k1601B}, convolution method \cite{Altarelli2001125, 2005PhRvD..72e5018B, 1992PhRvD..45..982K, 1992ZPhyC..56..273U, 1992PhRvD..46.2982U, 2000JPhG...26..387K}, decay-chain method \cite{Altarelli2001125}, and semi-analytical approach \cite{bardin1999standard}. All of these approximations get a strict analytical formulation within the framework of the factorization method. For instance, the narrow-width approximation includes five assumptions, which were considered in detail in \cite{2007PhRvL..99k1601B}, while the factorization method contains only one assumption---non-factorable corrections are small (the fifth assumption of the narrow-width approximation \cite{2007PhRvL..99k1601B}).

It is important to estimate the error of the calculations in the model with continuous masses that is the deviation of the model results from the standard ones. For a scalar UP the error equals zero. For a vector UP the error comes from the following difference:
\begin{equation}\label{4.6}
\eta_{\mu\nu}(q^2)-\eta_{\mu\nu}(m^2)
 = q_{\mu} q_{\nu} \frac{m^2-q^2}{m^2 q^2}.
\end{equation}
In the case of the meson-pair production $e^+e^-\to \rho^0,\omega, \phi \to\pi^+\pi^-, K^+K^-,\rho^+\rho^-,...$ the deviation equals to zero, due to the vanishing contribution of the transverse parts of the amplitudes in cases:
\begin{equation}\label{4.7}
\mathcal{M}^\mathrm{trans}(q)
\sim \bar{e}^-(\mathbf{p}_1)\nt{q}e^-(\mathbf{p}_2)
= \bar{e}^-(\mathbf{p}_1)(\nt{p}_1+\nt{p}_2)e^-(\mathbf{p}_2)=0.
\end{equation}
In the case of the high-energy collisions $e^+e^-\to Z^0 \to f\bar{f}$ (we neglect $\gamma Z$ interference) the transverse part of the amplitude is \cite{2009IJMPA..24.5765K}
\begin{equation}\label{4.8}
\mathcal{M}^\mathrm{trans}(q)
\sim \bar{e}^-(\mathbf{p}_1)\nt{q}(c_e-\gamma_5 d_e)e^-(\mathbf{p}_2)
\cdot \bar{f}^+(\mathbf{k}_1)(c_f-\gamma_5 d_f)f^+(\mathbf{k}_2).
\end{equation}
We then get at $q^2\approx M^2_Z$
\begin{equation}\label{4.9}
\delta \mathcal{M} \sim \frac{m_e m_f}{M^2_Z}\frac{M_Z-q}{M_Z}.
\end{equation}
Thus, the error of the factorization method at the vicinity of resonance is always small. Similar estimates can be easily made for the case of a spinor UP.

The relative deviation of the partial cross section for the boson-pair production with consequent decays of the bosons to fermion pairs is
\begin{equation}\label{4.10}
\epsilon_f \sim
4\frac{m_f}{M}\biggl[1-M\int_{m^2_f}^{s}\frac{\rho(q^2)}{q} dq^2 \biggr],
\end{equation}
where $M$ is a boson mass. If the fermion $f$ is a $\tau$ lepton, the deviation is maximal, $\epsilon_{\tau}\sim 10^{-3}$. It should be noted that the deviations which are caused by the approach at the tree level are significantly smaller than the errors related to the uncertainties in taking account of radiative corrections \cite{2010PAN....73.1622K}.

In the case of $\mu$ and $\tau$ decays the relative deviations of the widths from the standard results are as follows:
\begin{equation}\label{4.11}
\epsilon(\mu \to e\nu\bar\nu) \approx 5 \cdot 10^{-4},
\qquad
\epsilon(\tau \to e\nu\bar\nu) \approx 3 \cdot 10^{-6},
\qquad
\epsilon(\tau \to \mu\nu\bar\nu) \approx 3 \cdot 10^{-2}.
\end{equation}
The error in the last case is noticeable because of the factor $m_\mu/m_\tau$. In the case of a spinor UP the deviation is of the order of $(M_f-q)/M_f$. It could be substantial, if $q$ is far from the resonance region, but it is suppressed by averaging over variable mass as in Eq.~\eqref{4.10}. So, the accuracy of the existing experimental data on the processes with participation $Z$, $W$, and $t$ does not make it possible to distinguish the predictions of continuous-mass model from the standard ones.


\section{\label{sec:5}Conclusion}


The factorization method is a convenient analytical way to describe decays and scattering processes. It follows directly from the model of UP with a continuous mass that is based on the mass-lifetime uncertainty relation. The factorization simplifies calculations considerably and gives compact universal formulae for decay widths and cross sections of complicated processes. The method is phenomenological in its character and has been applied so far to the description of processes in Born approximation. Its applicability beyond tree level is an open question.

In this work, we have shown that in the model with continuous masses the exact factorization holds for time-like intermediate states of arbitrarily high spin tree diagrams. The factorization is due to the specific tensor-spinor structure of the propagators in the model. We have derived the factorization formulae for some particular cascade processes. The model of UP with continuous mass is an analytical analogue of the narrow-width approximation. This approach can be considered as a convenient approximation to the conventional field theory which is always valid in the vicinity of the resonance peak, if nonresonance contributions are negligible. To distinguish experimentally between predictions of the continuous-mass model and standard calculations, we need a considerable improvement in the accuracy of collider measurements. Note also that the method allows us to take into account all factorable corrections.


\appendix


\section{\label{app:a}Factorization of amplitude squared}


The normalization of the amplitudes can be defined by their relations to the observables. Differential width of the decay process $a(p)\to X_F(k_1,\dots k_f)$, where $X_F$ is a set of $f$ final particles with 4-momenta $k_i$, $i=1,\dots f$, has the form:
\begin{equation}\label{A2.1}
d\Gamma(p,k_1,\dots k_f)
= \frac{1}{2\pi K^2}\delta\biggl(p-\sum k_i\biggr)
|\mathcal{M}(p,k_1,\dots k_f)|^2 \prod d\mathbf{k}_i.
\end{equation}
where $K=(2\pi)^{3(s+f)/2+4(m-n)}$. Here, $s = 1$ and $f$ are numbers of initial and final states, $m$ and $n$ are numbers of internal lines and vertices.

Differential cross section of the process $a(p_a)b(p_b)\to X_F(k_1,\dots k_f)$ is:
\begin{equation}\label{A2.2}
d\sigma(p_a,p_b,k_1,\dots k_f)
=\frac{(2\pi)^2}{v(p)K^2} \delta\biggl(q-\sum k_i\biggr)
|\mathcal{M}(p_a,p_b,k_1,\dots k_f)|^2
\prod d\mathbf{k}_i,
\end{equation}
where $q=p_a+p_b$ and $v(p)p^0_ap^0_b=\sqrt{(p_a p_b)^2-m^2_am^2_b}$. The amplitude $\mathcal{M}(p,k)$ does not contain any $(2\pi)$ and is normalized by $(2p_0)^{-1/2}$ for the boson states. For the spinor states, the same normalizing coefficients in the amplitude squared come from the polarization matrices \eqref{2.2} and \eqref{2.2a}.

\subsection{UP with a spin $j_R=1/2$}

Let us consider a process $X_I \to R \to X_F$, where $R$ is a spin-$1/2$ UP with a large width. To reduce the number of nonessential indices in the following equations, we restrict ourselves to the case of a two-particle initial $X_I$ and multi-particle final state $X_F$ consisting of one spin-$1/2$ particle and a number of scalar particles (see Fig.~\ref{fig:Born4}). As it will be easily seen, however, all of the following results are valid for $X_I$ and $X_F$ including any number of particles of arbitrarily high spins. The vertices (1) and (2) in Fig.~\ref{fig:Born4} are not specified and can involve loops. The amplitude of this process is written as follows:
\begin{equation}\label{A2.3}
\mathcal{M}(p,q,k) = \frac{1}{\sqrt{2p^0_b}}\bar\psi^+_F(\mathbf{k}_1)
\Gamma_{(2)}(k,q) \frac{\nt{q}+q}{P_R(q)}
 \Gamma_{(1)}(p,q) \psi^-_I(\mathbf{p}_a)
\prod_{i=2}^f \frac{1}{\sqrt{2k^0_i}}.
\end{equation}
Here, $p=p_a,p_b$ and $k=k_1,k_2,\dots k_f$ are momenta of the initial and final particles, respectively. Hermitian conjugation of the amplitude $\mathcal{M}^\dagger(p,q,k)$ is taken so that equalities $(\bar\psi^+_1 \Gamma \psi^-_2)^\dagger =\bar\psi^+_2 \bar\Gamma \psi^-_1$ and $\bar\Gamma = \gamma_0 \Gamma^+ \gamma_0$ hold. The amplitude squared $|\mathcal{M}(p,q,k)|^2$ (summed over the spin states of the final particles and averaged over the spin states of the initial ones) is as follows:
\begin{equation}\label{A2.6}
|\mathcal{M}(p,q,k)|^2 =
\frac{(2p^0_a 2p^0_b \prod 2k^0_i)^{-1}}{(2j_a+1)(2j_b+1)} \frac{1}{|P_R(q)|^2}
\Tr\Bigl[(\nt{p}_a+m_a)\bar\Gamma_{(1)}(\nt{q}+q)\bar\Gamma_{(2)}
(\nt{k}_1+m'_1)\Gamma_{(2)}(\nt{q}+q)\Gamma_{(1)}\Bigr].
\end{equation}
In Eq.~\eqref{A2.6}, the spins of the initial particles $j_a=1/2$ and $j_b=0$ are preserved as symbols for generality.

We can also write the amplitudes of the decays $\psi_R \to X_I$ and $\psi_R \to X_F$
\begin{equation}\label{A2.7}
\begin{aligned}
\mathcal{M}_1(p,q)
&=\mathcal{M}(\psi_R(q) \to X_I)
= \frac{1}{\sqrt{2p^0_b}}
\bar\psi^+_I(\mathbf{p}_a)\bar\Gamma_{(1)}(p,q)\psi^-_R(\mathbf{q}),
\\
\mathcal{M}_2(k,q)
&= \mathcal{M}(\psi_R(q) \to X_F)
= \prod_{i=2}^f \frac{1}{\sqrt{2k^0_i}} \cdot
\bar\psi^+_F(\mathbf{k}_1)\Gamma_{(2)}(k,q)\psi^-_R(\mathbf{q}).
\end{aligned}
\end{equation}
The amplitudes squared $|\mathcal{M}_{1,2}|^2$ are as follows:
\begin{equation}\label{A2.8}
\begin{aligned}
|\mathcal{M}_1(p,q)|^2 = {}& \Tr \hat{M}_1(p,q)
= \frac1{8 L_R p^0_a p^0_b q^0}
\Tr\bigl[(\nt{q}+q)\Gamma_{(1)}(\nt{p}_a+m_a)\bar\Gamma_{(1)}\bigr];
\\
|\mathcal{M}_2(k,q)|^2 = {}& \Tr\hat{M}_2(k,q)
= \frac1{L_R (2q^0) \prod (2k^0_i)}
\Tr\bigl[(\nt{q}+q)\bar\Gamma_{(2)}(\nt{k}_1+m'_1) \Gamma_{(2)} \bigr],
\end{aligned}
\end{equation}
where $L_R = 2J_R + 1$. The above equations define $4\times4$ matrices $\hat{M}_{1}(p,q)$ and $\hat{M}_{2}(k,q)$ that have a structure $\hat{M}_{1,2}=\hat{M}'_{1,2}(\nt{q}+q)$.

Now we can prove the following identity:
\begin{align}\label{A2.9}
J\bigl[|\mathcal{M}(p,q,k)|^2\bigr]
&{}= \frac{L_R^2 (2q^0)^2}{L_a L_b|P_R(q)|^2}
\Tr J\bigl[ \hat{M}_2(k,q) \hat{M}_1(p,q) \bigr]
\nonumber\\
&{} = \frac{L_R (2q^0)^2}{L_a L_b|P_R(q)|^2}
\Tr\hat{M}_1(k,q) \cdot J \bigl[ \Tr \hat{M}_2(p,q) \bigr].
\end{align}
Here, $J[X]$ stands for the integration over the phase space of the final state $X_F$
\begin{equation}\label{app:int1}
J[X] = \int X \delta\Bigl(q-\sum k_i \Bigr) \prod d\mathbf{k}_i.
\end{equation}
To this end, let us consider the trace in the right-hand side of Eq.~\eqref{A2.9}, $T(p,q) \equiv \Tr J \bigl[\hat{M}_2(k,q)
\hat{M}_1(p,q)\bigr]$. Since $(\nt{q}+q)^2 = 2q(\nt{q}+q)$, we obtain
\begin{equation}\label{A2.10}
T(p,q) = \frac{1}{4q^2}\Tr J\bigl[ \hat{M}_2(k,q)(\nt{q}+q)\hat{M}_1(p,q)(\nt{q}+q)\bigr]
= \frac{1}{4q^2}\Tr\bigl[\hat{A}(q) \hat{M}_1(p,q)\bigr],
\end{equation}
where $\hat{A}(q) = ({\nt{q}}+q) J\bigl[\hat{M}_2(k,q)\bigr] (\nt{q}+q)$ and the following identity holds:
\begin{equation}\label{A}
    \hat{A}(q) = \frac1{4q^2} ({\nt{q}}+q) \hat{A}(q) (\nt{q}+q).
\end{equation}
The matrix $\hat{A}(q)$ can be expanded in terms of linearly independent matrices as follows:
\begin{equation}\label{A2.12}
\hat{A}(q)= ({\nt{q}}+q) (a_1+b_1\gamma_5)+ ({\nt{q}}-q) (a_2+b_2\gamma_5).
\end{equation}
Here $a_i = a_i(q)$ and $b_i = b_i(q)$ are scalar coefficients. Substituting the decomposition \eqref{A2.12} into the righthand side of Eq.~\eqref{A}, we obtain $\hat{A}(q)= a_1 ({\nt{q}}+q)$. As a result, we have
\begin{equation}\label{A2.14}
\hat{A}(q)=q(\nt{q}+q)\Tr J \bigl[\hat{M}_2(k,q)\bigr]
\end{equation}
and consequently
\begin{equation}\label{A2.15}
T(p,q) = \frac12 \Tr J \bigl[\hat{M}_2(k,q)\bigr] \cdot \Tr\hat{M}_1(p,q),
\end{equation}
where we have used the structure $\hat{M}_1(p,q)=\hat{M}'_1(p,q)(\nt{q}+q)$. Finally, we have proved the identity \eqref{A2.9} and the factorization of the amplitude squared for the case of a spin-$1/2$ UP ($2j_R+1=2$):
\begin{equation}\label{A2.16}
J\bigl[|\mathcal{M}(p,q,k)|^2\bigr]=
\frac{L_R (2q_0)^2}{L_a L_b|P_R(q)|^2}
|\mathcal{M}_1(p,q)|^2 J\bigl[|\mathcal{M}_2(k,q)|^2\bigr].
\end{equation}

\subsection{\label{app:spin1}UP with a spin $j_R=1$}

Now, we proceed to a process $X_I \to V_R \to X_F$ with a vector UP in the intermediate state. In the case of scalar initial and final particles, the amplitude is
\begin{equation}\label{A2.17}
\mathcal{M}(p,q,k)
= \frac{1}{\sqrt{2p^0_a}} \frac{1}{\sqrt{2p^0_b}} \prod_{i=1}^f \frac1{\sqrt{2k^0_i}}
\cdot
\Gamma^{(1)}_{\mu}(p,q) \frac{\eta^{\mu\nu}(q)}{P_R(q)} \Gamma^{(2)}_{\nu}(k,q).
\end{equation}
Here, $\Gamma^{(1,2)}_\mu$ are vertices; $\eta_{\mu\nu}(q) = -g_{\mu\nu} + q_\mu q_\nu/q^2$. For the amplitude squared we then have
\begin{equation}\label{A2.18}
|\mathcal{M}(p,q,k)|^2=
\frac{[4 p^0_a p^0_b \prod(2k^0_i)]^{-1}}{(2j_a+1)(2j_b+1)} \frac{1}{|P_R(q)|^2}
\Gamma^{(1)*}_{\mu'}(p,q) \Gamma^{(1)}_\mu(p,q) \eta^{\mu\nu}(q)
\eta^{\mu'\nu'}(q) \Gamma^{(2)*}_{\nu'}(q,k) \Gamma^{(2)}_{\nu}(q,k).
\end{equation}
For the processes $V_R \to X_I$ and $V_R \to X_F$ the amplitudes and their squares can be written as
\begin{align}\label{A2.19}
\mathcal{M}_1(p,q) &{}=
\frac{1}{\sqrt{2q^0}} \frac{1}{\sqrt{2p^0_a}} \frac{1}{\sqrt{2p^0_b}} e^\mu(\mathbf{q}) \Gamma^{(1)}_\mu(p,q),
\\
\mathcal{M}_2(k,q)
&{}=\frac{1}{\sqrt{2q^0}} \prod_{i=1}^f \frac1{\sqrt{2k^0_i}} \cdot
e^\nu(\mathbf{q}) \Gamma^{(2)}_\nu(k,q)
\end{align}
and
\begin{align}\label{A2.21}
|\mathcal{M}_1(p,q)|^2 ={}&
\frac{(8q^0p^0_a p^0_b)^{-1}}{2j_R+1}
\Gamma^{(1)*}_{\mu'}(p,q) \Gamma^{(1)}_{\mu}(p,q)\eta^{\mu\mu'}(q),
\\ \label{A2.21b}
|\mathcal{M}_2(k,q)|^2 ={}&
\frac{[2q^0 \prod (2k^0_i)]^{-1}}{2j_R+1}
\Gamma^{(2)*}_{\nu'}(q,k) \Gamma^{(2)}_{\nu}(q,k) \eta^{\nu\nu'}(q).
\end{align}
For brevity we introduce tensors $T^{(1)}_{\mu\mu'}(p,q)$ and $T^{(2)}_{\nu\nu'}(k,q)$ so that Eqs.~\eqref{A2.18}, \eqref{A2.21}, and \eqref{A2.21b} take the form
\begin{align}\label{A2.23}
|\mathcal{M}(p,k,q)|^2
 = {}& \frac{(2j_R+1)^2(2q^0)^2}{(2j_a+1)(2j_b+1)|P_R(q)|^2}
  T^{(1)}_{\mu\mu'}(p,q) \eta^{\mu\nu}(q) \eta^{\mu'\nu'}(q) T^{(2)}_{\nu\nu'}(k,q),
\\ \label{A2.22}
|\mathcal{M}_1(p,q)|^2 = {}& T^{(1)}_{\mu\mu'}(p,q) \eta^{\mu\mu'}(q),
\\
|\mathcal{M}_2(k,q)|^2 = {}& T^{(2)}_{\nu\nu'}(k,q) \eta^{\nu\nu'}(q).
\end{align}
We can show that the exact factorization arises as a result of integrating over the phase space of the final state $X_F$
\begin{equation}\label{A2.24}
J\bigl[|\mathcal{M}(p,k,q)|^2\bigr] =
\frac{(2j_R+1)^2(2q^0)^2}{(2j_a+1)(2j_b+1)|P_R(q)|^2}
T^{(1)}_{\mu\mu'}(p,q) \eta^{\mu\nu}(q) \eta^{\mu'\nu'}(q)
J\bigl[T^{(2)}_{\nu\nu'}(k,q) \bigr].
\end{equation}
The tensor integral $J[T^{(2)}_{\nu\nu'}(k,q)]$ can be decomposed in the following way:
\begin{equation}\label{A2.25}
J\bigl[T^{(2)}_{\nu\nu'}(k,q)\bigr]
= \frac{1}{2j_R+1} \bigl[T(q)\eta_{\nu\nu'}(q) + S(q)q_{\nu}q_{\nu'}\bigr].
\end{equation}
The second term does not contribute to \eqref{A2.24} because of the property $\eta^{\mu\nu}(q) q_\nu=0$. Multiplying Eq.~\eqref{A2.25} by $\eta^{\nu\nu'}(q)$, we find
\begin{equation}\label{A2.26}
T(q) = J\bigr[T^{(2)}_{\nu\nu'}(k,q)\eta^{\nu\nu'}(q)\bigr]
= J\bigl[|\mathcal{M}_2(k,q)|^2\bigr].
\end{equation}
Finally, substituting Eqs.~\eqref{A2.25} and \eqref{A2.26} into Eq.~\eqref{A2.24} gives
\begin{equation}\label{A2.27}
J(|\mathcal{M}(p,k,q)|^2)=
 \frac{(2j_R+1)(2q^0)^2}{(2j_a+1)(2j_b+1)|P_R(q)|^2}
 |\mathcal{M}_1(p,q)|^2 J\bigl[|\mathcal{M}_2(k,q)|^2\bigr].
\end{equation}

Therefore, we have proved that being integrated over the final momenta the transition probability of a process with a vector UP in the intermediate time-like state is factorized exactly in the framework of the model with the propagator \eqref{2.1}.

\subsection{\label{app:a3}UP with a spin $j_R \geqslant 3/2$}

It is also necessary to discuss the case of UP carrying arbitrarily high spin $j_R$. In the model of UP with continuous mass, the tensor-spinor structure of the propagators of higher-spin particles is given by the projection operators introduced in \cite{1957PhRv..106..345B, springerlink:10.1007/BF02747684}. For integer spins $j_R = \ell = 1, 2,\dots$, the simplest projector $\ell = 1$ is well known:
\begin{equation}\label{A2.28}
P^{(1)}_{\mu\nu}(q)=g_{\mu\nu}-\frac{q_{\mu}q_{\nu}}{q^2}=-\eta_{\mu\nu}(q).
\end{equation}
The projection operators for higher integer spins $\ell \geqslant 2$ are defined by the recurrence relation
\begin{equation}\label{A2.29}
P^{(\ell)}_{\bar{\mu}\bar{\nu}}(q)
=
\frac{1}{\ell^2}
\biggl\{ P^{(\ell-1)}_{\bar{\mu}^f\bar{\nu}^g}(q) P^{(1)}_{\mu_f\nu_g}(q)
- \frac{1}{2(2\ell-1)} \Bigl[P^{(\ell-1)}_{\bar{\mu}^f\mu_f\bar{\nu}^{gh}}(q)P^{(1)}_{\nu_{gh}}(q)
+P^{(\ell-1)}_{\bar{\nu}^f\nu_f\bar{\mu}^{gh}}(q)P^{(1)}_{\mu_{gh}}(q)\Bigr]\biggr\},
\end{equation}
where $\bar{\mu}=\mu_1\mu_2...\mu_\ell$, $\bar{\mu}^f=\mu_1\mu_2...\mu_{f-1}\mu_{f+1}...\mu_{\ell}$, $\mu_{fg}=\mu_f\mu_g$ are multi-indices. The summation over repeated Latin indices $f, g, h$ of the Greek multi-indices is implied from 1 to $\ell$.

The projectors possess the following properties \cite{1957PhRv..106..345B, springerlink:10.1007/BF02747684}
\begin{equation}\label{A2.30}
\begin{gathered}
    P^{(\ell)}_{\mu_1\cdots\mu_a\mu_b\cdots\mu_\ell\bar\nu} =
    P^{(\ell)}_{\mu_1\cdots\mu_b\mu_a\cdots\mu_\ell\bar\nu},
    \qquad
    P^{(\ell)}_{\bar\mu\bar\nu} = P^{(\ell)}_{\bar\nu\bar\mu},
\\
    P^{(\ell)}_{\bar\mu\bar\lambda} P^{(\ell)}{}^{\bar\lambda}_{\bar\nu}
    =P^{(\ell)}_{\bar\mu\bar\nu},
    \qquad
    P^{(\ell)}{}^{\bar\lambda}_{\bar\lambda} = 2\ell+1,
\\
    q^{\mu_a} P^{(\ell)}_{\bar\mu\bar\nu} = 0,
    \qquad
    g^{\mu_a\mu_b} P^{(\ell)}_{\bar\mu\bar\nu} = 0
\end{gathered}
\end{equation}
for any $a, b = 1,\dots \ell$. For $\ell=1,2,3,4$ the relations \eqref{A2.30} are checked straightforwardly. For $\ell \geqslant 4$ these properties can be proved by induction.

The projectors for half-integer spins $j_R = \ell + 1/2 = 3/2, 5/2,\dots$ are written as
\begin{equation}\label{2.17a}
P^{(\ell+\frac12)}_{\bar{\mu}\bar{\nu}}(q)
 = \frac{\ell+1}{2\ell+3}\gamma^{\alpha}\gamma^{\beta}
P^{(\ell+1)}_{\bar{\mu}\alpha\bar{\nu}\beta}(q),
\qquad
\ell \geqslant 1.
\end{equation}
For $\ell = 1$ and $\ell = 2$ the projectors \eqref{2.17a} coincide
with the known expressions for $P^{(3/2)}_{\mu\nu}$
\cite{2004pilling, 2005pilling} and
$P^{(5/2)}_{\mu_1\mu_2\nu_1\nu_2}$ \cite{1979berends}.

The properties of the projectors \eqref{2.17a} are
\begin{equation}\label{2.18b}
\begin{gathered}
P^{(\ell+\frac12)}_{\bar{\mu}\bar{\lambda}}P^{\bar{\lambda}}_{(\ell+\frac12)\bar{\nu}}=P^{(\ell+\frac12)}_{\bar{\mu}\bar{\nu}},
\qquad
\bar P^{(\ell+\frac12)}_{\bar{\mu}\bar{\nu}}(q) = \bar P^{(\ell+\frac12)}_{\bar{\nu}\bar{\mu}}(q),
\\
\Tr P^{(\ell+\frac12)\bar{\mu}}_{\bar{\mu}} =4(\ell+1),
\qquad
\nt{q}P^{(\ell+\frac12)}_{\bar{\mu}\bar{\nu}}(q)=P^{(\ell+\frac12)}_{\bar{\mu}\bar{\nu}}(q)\nt{q},
\\
q^{\mu_a}P^{(\ell+\frac12)}_{\bar{\mu}\bar{\nu}}(q)=0,
\qquad
\gamma^{\mu_a}P^{(\ell+\frac12)}_{\bar{\mu}\bar{\nu}}(q)=0.
\end{gathered}
\end{equation}
These identities follow from the identities \eqref{A2.30} for the projectors $P^{(\ell)}_{\bar\mu\bar\nu}$. It is the properties \eqref{A2.30} and \eqref{2.18b} which allow us to generalize the consideration of the previous subsections to the case of arbitrary spins.

Now, it is easy to see that the factorized formula \eqref{A2.27} for the amplitude squared is valid for any integer spin of UP. Indeed, in the proof of Appendix \hyperref[app:spin1]{A.2} we just need to change all Lorentz indices to multi-indicies ($\mu \to \bar\mu$, $\nu \to \bar\nu$, etc.) and spin-1 projectors to spin-$\ell$ ones ($\eta_{\mu\nu} \to -P^{(\ell)}_{\bar\mu\bar\nu}$). By similar substitutions we can generalize the proof of the relation \eqref{A2.27} to arbitrarily high half-integer spins.


\section{\label{app:b}Factorized formula for cross section and convolution
formula for decay width}


In this Appendix we derive universal factorized formula for the cross section of the scattering $ab \to R \to X_F$ and convolution formula for the decay $a \to b R \to b X_F$.

Using Eqs.~\eqref{A2.1}, \eqref{A2.2}, and \eqref{A2.27}, we find the cross section of the process under consideration
\begin{align}\label{A3.1}
\sigma(p_1,p_2) &{}=
\frac{(2\pi)^2}{K^2 v(p)} J \bigl[ |\mathcal{M}(p,k)|^2 \bigr]
\nonumber\\
&{}=\frac{(2j_R+1)(2q^0)^2 |\mathcal{M}_1(p,q)|^2 J\bigl[ |\mathcal{M}_2(k,q)|^2 \bigr]}{(2\pi)K_2^2(2j_a+1)(2j_b+1)v(p)|P_R(q)|^2}
\nonumber\\
&{}=\frac{(2j_R+1)(2q^0)^2|\mathcal{M}_1(p,q)|^2\Gamma(R(q) \to X_F)}{(2j_a+1)(2j_b+1)v(p)|P_R(q)|^2},
\end{align}
where $K=(2\pi)^{3f/2-1}=2\pi K_1K_2$ and $K_1=(2\pi)^{1/2}$. The amplitude squared $|\mathcal{M}_1(p,q)|^2$ is related to the decay width
\begin{equation}\label{A3.2}
\Gamma(R(q) \to ab)
=\frac{1}{(2\pi)^2}\int\delta(q^0-p^0_a-p^0_b) |\mathcal{M}_1(p,q)|^2 d\mathbf{p}_a
\nonumber\\
=\frac{1}{\pi}|\mathcal{M}^{(1)}(p,q)|^2 p^0_a p^0_b \frac{p}{q^0}.
\end{equation}
In the centre-of-mass frame, it follows from Eq.~\eqref{A3.2} that
\begin{equation}\label{A3.3}
|\mathcal{M}_1(p,q)|^2=2\pi \frac{\Gamma(R(q)\to
a_1a_2)}{p^0_a p^0_b\bar{\lambda}(m_1^2,m_2^2;q^2)},
\end{equation}
where the function $\bar\lambda(m^2_a,m^2_b;q^2) = 2v(p)p^0_a p^0_b/q^2$ is defined by Eq.~\eqref{Kallen}.

Upon substituting Eq.~\eqref{A3.3} into Eq.~\eqref{A3.1} we get
\begin{equation}
\sigma(q^2)=
\frac{16\pi(2j_R+1)}{(2j_a+1)(2j_b+1)\bar{\lambda}^2(m^2_a,m^2_b;q^2)|P_R(q^2)|^2}
\Gamma(R(q)\to ab) \Gamma(R(q)\to X_F).\label{A3.4}
\end{equation}
Universal factorized formula \eqref{A3.4} describes cross section of the process $ab\to R\to X_F$, where $R$ is UP with arbitrarily high spin in the time-like intermediate state, $X_F$ is any set of final particles, and the vertices $\Gamma_{(1,2)}$ have an arbitrary (loop) structure. This result is a generalization of the corresponding formula for the tree processes $ab\to R\to cd$ that was obtained for some particular interactions mediated by scalar, vector, and spinor UP $R$ in \cite{2008IJMPA..23.4509K}.

Now, let us consider the decay $a\to bR\to bX_F$ with all particles and vertices being of any type. The factorized relation for the amplitude squared of the decay differs from that for the scattering amplitude \eqref{A2.27} by a numeric factor
\begin{equation}\label{A3.5}
J\bigl[|\mathcal{M}(p,k,q)|^2\bigr]
= \frac{(2q^0)^2}{|P_R(q)|^2}
|\mathcal{M}_1(p,q)|^2 J \bigl[|\mathcal{M}_2(k,q)|^2\bigr],
\end{equation}
In Eq.~\eqref{A3.5} the amplitudes $\mathcal{M}_1(p,q)$ and $\mathcal{M}_2(k,q)$ correspond to the subprocesses $a\to bR$ and $R \to X_F$, respectively. In terms of the normalized amplitudes $\mathcal{A}_i=\mathcal{M}_i/K_i$, the decay width \eqref{A2.1} is
\begin{align}\label{A3.6}
\Gamma &{}= \frac{1}{(2\pi)^3}\int
\frac{(2q^0)^2}{|P_R(q)|^2}|\mathcal{A}_1(p,q)|^2
|\mathcal{A}_2(k,q)|^2
\delta(p-k_1-q) \prod_{i=1}^{f} d\mathbf{k}_i
\nonumber\\
&{}= \frac{4}{(2\pi)^2}\int \frac{(q^0)^2}{|P_R(q)|^2}|\mathcal{A}_1(p_1,q)|^2
 \Gamma(R(q)\to X_F) d\mathbf{k}_1,
\end{align}
where $q = k_2 + k_3 + \cdots + k_f$. Then, we find from Eq.~\eqref{A3.3} that
\begin{equation}\label{A3.8}
\Gamma(a\to bR(q))=2|\mathcal{A}_1(p,q)|^2|\mathbf{k}_1|k^0_1\frac{q^0}{p^0},
\end{equation}
where $\mathcal{A}_1 = \mathcal{M}_1/\sqrt{2\pi}$. Substituting \eqref{A3.8} into \eqref{A3.6} and noting that $d\mathbf{k}_1 = 2\pi dq^2 |\mathbf{k}_1|k^0_1/p^0$ in the rest frame of the decayed particle $a$, we get
\begin{equation}\label{A3.9}
\Gamma(a\to bR\to bX_F)
=\int \Gamma(a\to bR(q)) \frac{q\Gamma(R(q)\to X_F)}{\pi|P_R(q)|^2}dq^2.
\end{equation}
Here, the limits of integration are determined by the kinematics of the process and we have used the relation $q\Gamma(q) = q^0\Gamma(q^0)$, where $\Gamma(q^0)$ is the width in the centre-of-mass frame, $\mathbf{q}=0$. Summation over all decay channels of UP $R(q)$ gives:
\begin{equation}\label{A3.10}
\Gamma(a\to bR)=\int \Gamma(a\to bR(q)) \rho_R(q)dq^2,
\end{equation}
where $\rho_R(q)=q\Gamma^\mathrm{tot}R(q)/(\pi|P_R(q)|^2)$ is interpreted in the framework of the model of UP with continuous mass as the probability density of the mass parameter $m^2=q^2$.


\end{document}